\documentclass[pra,reprint,superscriptaddress,bibnotes]{revtex4-1}

\usepackage[T1]{fontenc}
\usepackage[utf8]{inputenc}
\usepackage[australian]{babel}
\usepackage[a4paper,centering,hmargin=1.75cm,vmargin=2cm]{geometry}
\usepackage{amsmath,amssymb,bm}
\usepackage{graphicx,xcolor}
\usepackage{titlesec}
\usepackage{microtype}
\usepackage{braket,siunitx}
\usepackage[version=3]{mhchem}
\usepackage{bbold}
\usepackage[normalem]{ulem}

\titleformat*{\paragraph}{\itshape}

\newcommand{\LS}[2]{{}^2{\textrm{#1}}_{#2}}

\begin{document}

\title{Predicting molecular vibronic spectra using time-domain analog quantum simulation}

\affiliation{School of Chemistry, University of Sydney, NSW 2006, Australia}
\affiliation{School of Physics, University of Sydney, NSW 2006, Australia}
\affiliation{ARC Centre of Excellence for Engineered Quantum Systems, University of Sydney, NSW 2006, Australia}
\affiliation{University of Sydney Nano Institute, University of Sydney, NSW 2006, Australia}

\author{Ryan J. MacDonell}
\thanks{These authors contributed equally to this work}
\affiliation{School of Chemistry, University of Sydney, NSW 2006, Australia}
\affiliation{ARC Centre of Excellence for Engineered Quantum Systems, University of Sydney, NSW 2006, Australia}
\affiliation{University of Sydney Nano Institute, University of Sydney, NSW 2006, Australia}

\author{Tomas Navickas}
\thanks{These authors contributed equally to this work}
\affiliation{School of Physics, University of Sydney, NSW 2006, Australia}
\affiliation{ARC Centre of Excellence for Engineered Quantum Systems, University of Sydney, NSW 2006, Australia}

\author{Tim F. Wohlers-Reichel}
\affiliation{School of Physics, University of Sydney, NSW 2006, Australia}
\affiliation{ARC Centre of Excellence for Engineered Quantum Systems, University of Sydney, NSW 2006, Australia}

\author{Christophe~H.~Valahu}
\affiliation{School of Physics, University of Sydney, NSW 2006, Australia}
\affiliation{ARC Centre of Excellence for Engineered Quantum Systems, University of Sydney, NSW 2006, Australia}

\author{Arjun D. Rao}
\affiliation{School of Physics, University of Sydney, NSW 2006, Australia}
\affiliation{ARC Centre of Excellence for Engineered Quantum Systems, University of Sydney, NSW 2006, Australia}

\author{Maverick J. Millican}
\affiliation{School of Physics, University of Sydney, NSW 2006, Australia}
\affiliation{ARC Centre of Excellence for Engineered Quantum Systems, University of Sydney, NSW 2006, Australia}

\author{Michael A. Currington}
\affiliation{School of Chemistry, University of Sydney, NSW 2006, Australia}

\author{Michael~J.~Biercuk}
\affiliation{School of Physics, University of Sydney, NSW 2006, Australia}
\affiliation{ARC Centre of Excellence for Engineered Quantum Systems, University of Sydney, NSW 2006, Australia}

\author{Ting Rei Tan}
\affiliation{School of Physics, University of Sydney, NSW 2006, Australia}
\affiliation{ARC Centre of Excellence for Engineered Quantum Systems, University of Sydney, NSW 2006, Australia}

\author{Cornelius Hempel}
\email{cornelius.hempel@psi.ch}
\affiliation{School of Physics, University of Sydney, NSW 2006, Australia}
\affiliation{ARC Centre of Excellence for Engineered Quantum Systems, University of Sydney, NSW 2006, Australia}
\affiliation{ETH Zurich-PSI Quantum Computing Hub, Laboratory for Nano and Quantum Technologies (LNQ), Paul Scherrer Institut, 5232 Villigen, Switzerland}

\author{Ivan Kassal}
\email{ivan.kassal@sydney.edu.au}
\affiliation{School of Chemistry, University of Sydney, NSW 2006, Australia}
\affiliation{ARC Centre of Excellence for Engineered Quantum Systems, University of Sydney, NSW 2006, Australia}
\affiliation{University of Sydney Nano Institute, University of Sydney, NSW 2006, Australia}

\begin{abstract}
Spectroscopy is one of the most accurate probes of the molecular world.
However, predicting molecular spectra accurately is computationally difficult because of the presence of entanglement between electronic and nuclear degrees of freedom. 
Although quantum computers promise to reduce this computational cost, existing quantum approaches rely on combining signals from individual eigenstates, an approach whose cost grows exponentially with molecule size.
Here, we introduce a method for scalable analog quantum simulation of molecular spectroscopy: by performing simulations in the time domain, the number of required measurements depends on the desired spectral range and resolution, not molecular size.
Our approach can treat more complicated molecular models than previous ones, requires fewer approximations, and can be extended to open quantum systems with minimal overhead. We present a direct mapping of the underlying problem of time-domain simulation of molecular spectra to the degrees of freedom and control fields available in a trapped-ion quantum simulator. We experimentally demonstrate our algorithm on a trapped-ion device, exploiting both intrinsic electronic and motional degrees of freedom, showing excellent quantitative agreement for a single-mode vibronic photoelectron spectrum of \ce{SO2}.
\end{abstract}

\maketitle

Spectroscopy---the measurement of light's interaction with matter---is one of the most important and precise experimental techniques for probing microscopic phenomena. 
The prediction of spectra via computational techniques serves as a benchmark for theoretical models of molecules, and good agreement between theory and experiment is essential if we are to truly understand chemical phenomena.

However, predicting molecular spectra remains difficult, especially for large molecules, those with significant entanglement between degrees of freedom, those that are open to an environment, or when high precision is required~\cite{domcke04}.
In particular, there are regimes where all common approximations break down, leaving critical cases without practical computational solutions.
For example, the Franck-Condon approximation of vibronic (vibrational + electronic) spectroscopy states that a transition is proportional to the overlap of initial and final vibrational wavefunctions~\cite{franck26,condon26}; while often a good approximation, it can fail when the dipole moment depends on nuclear displacements.
More generally, strong vibronic coupling between electronic states can lead to failures of the Born-Oppenheimer approximation and substantial nuclear-electronic entanglement~\cite{domcke04}.
Methods that include vibronic coupling are generally limited in molecule size or accuracy; for example, surface hopping uses an approximate form of the wavefunction and its evolution to reduce computational cost~\cite{tully90}, whereas multiconfigurational time-dependent Hartree is numerically exact but heuristic, with an unpredictable computational cost that can be exponential in system size~\cite{domcke04,worth08}.

Quantum computers promise to reduce the computational cost associated with predicting molecular spectra by offering a new means of computational simulation. As in other applications of quantum computers to chemistry~\cite{lamata14,peruzzo14,McClean.2016,arguelloluengo19,dipaolo20,schlawin21,seetharam21,young21,wang22,richerme22}, the advantage of quantum simulation lies in the ability to naturally represent complicated, entangled wavefunctions using quantum coherent degrees of freedom.
Indeed, recent proposals and experiments have shown that quantum computers can predict vibronic spectra~\cite{huh15,huh17,hu18,shen18,sawaya19,wang20,jnane21}, starting with the simulation of Franck-Condon spectra by encoding Duschinsky transformations in the displacement, squeezing, and unitary rotations of optical modes~\cite{huh15}.
This approach has been extended to thermal initial states~\cite{huh17} and non-Condon transitions~\cite{jnane21}, and demonstrated on other quantum platforms with experimentally accessible bosonic modes, including trapped ions~\cite{shen18} and circuit quantum electrodynamics (cQED)~\cite{wang20}.
These approaches are examples of analog quantum simulations, where the Hamiltonian of a system of interest is mapped onto a controllable quantum system in a laboratory.
In contrast, a digital quantum simulation (i.e., comprised of qubits and quantum gates) has also been proposed with a straightforward extension to include anharmonic vibrational modes~\cite{sawaya19}.

All existing approaches to quantum simulation for molecular spectroscopy suffer the same drawback, requiring exponential resources as the number of molecular vibrational modes increases. Most methods directly simulate the intensity of every spectral line~\cite{huh15,huh17,shen18,sawaya19,wang20,jnane21}, whose number can reach astronomical sizes even in small molecules. For example, if 10 states in each quantised vibrational mode are optically accessible, the number of states in an $N$-atom molecule is $10^{3N-6}$. Most approaches also rely on the Franck-Condon principle~\cite{huh15,huh17,shen18,sawaya19,wang20} and exclude vibronic coupling, which limits their use to describing single electronic states that are energetically separated from other electronic states.
The approach of Hu et~al.~\cite{hu18} predicts the spectrum of uncoupled, displaced harmonic oscillators by measuring overlaps between initial and final states; however, extending the method to more general chemical Hamiltonians requires exponential classical resources to predict the final state.

\begin{figure*}
    \centering
    \includegraphics[width=0.95\textwidth]{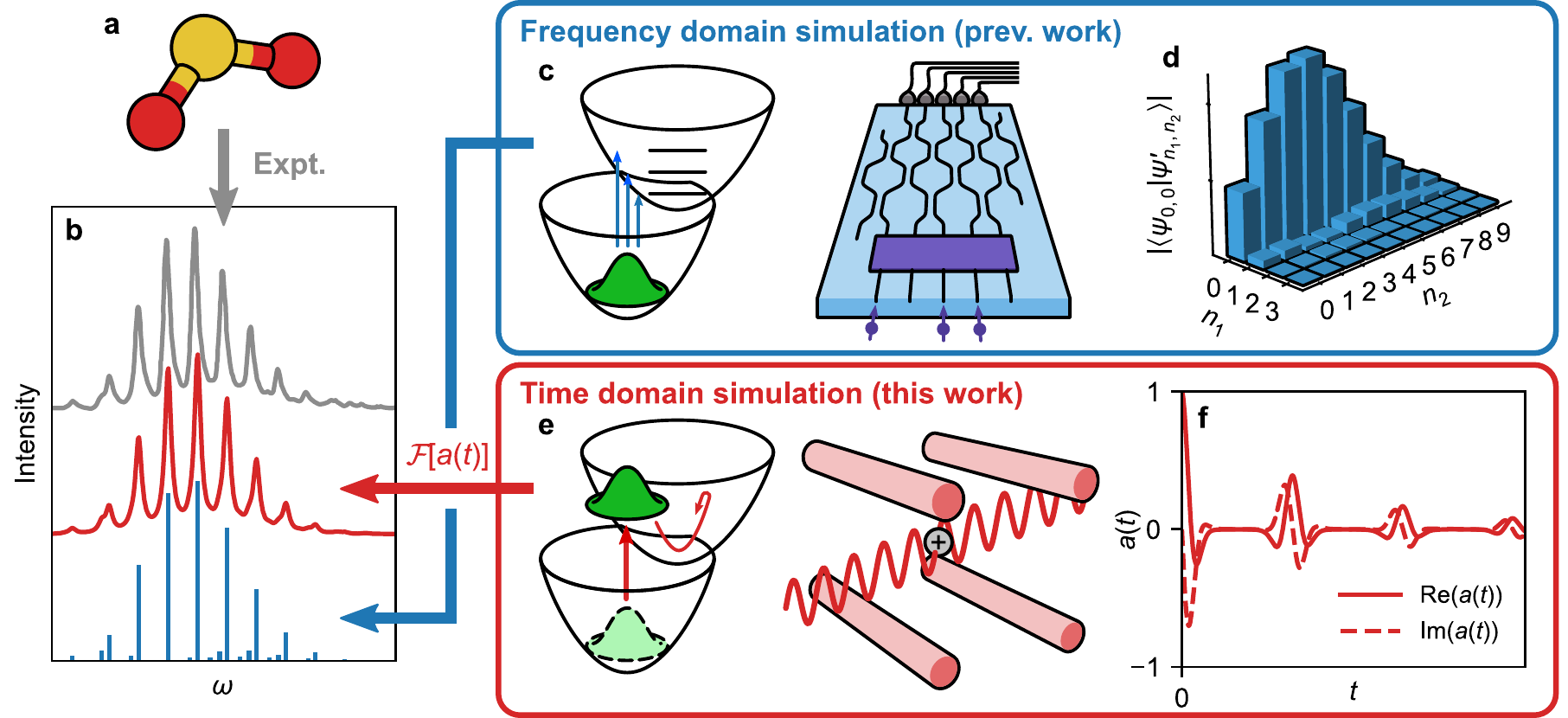}
    \caption{Different approaches to obtain a molecular vibronic spectrum. 
    \textbf{a}, Absorption of light by a molecule (shown: \ce{SO2}) can be measured experimentally to give a spectrum (grey line in \textbf{b} \cite{holland94}). 
    \textbf{c}, Several quantum simulation techniques can find Franck-Condon factors~\cite{huh15,huh17,shen18,wang20,jnane21} (illustrated with a boson-sampling optical circuit).
    \textbf{d}, These techniques measure the overlaps of the initial state with final eigenstates, here given by vibrational quanta $n_1$ and $n_2$ for two vibrational modes. The corresponding intensities give peak heights in the frequency domain (blue sticks in b).
    \textbf{e}, We show that the molecule can instead be mapped to a time-domain MQB simulation (illustrated with a trapped-ion simulator), with the ability to include vibronic coupling, mixed initial states, and open quantum systems~\cite{macdonell21}. The time-domain procedure is scalable with the number of vibrational modes, because it reconstructs the spectrum directly, not via exponentially many vibronic eigenstates.
    \textbf{f}, Measurements of the MQB simulation give the autocorrelation function $a(t)$, whose Fourier transform, $\mathcal{F}[a(t)]$, is the vibronic spectrum (red line in b), including lineshapes in the presence of noise.
    }
    \label{fig:map}
\end{figure*}

Systems with coupled electronic states and vibrational modes are well-suited for simulation on a class of analog quantum devices known as mixed qudit-boson (MQB) simulators~\cite{macdonell21}. These are comprised of a qudit, i.e., a $d$-level system with controllable transitions between all $d$ levels, and a set of quantum oscillators or resonators making up the bosonic modes. Example architectures include trapped ions and cQED. In addition to spectroscopic prediction~\cite{shen18,hu18,wang20,wang22}, MQB devices have been proposed for the analog quantum simulation of vibrationally assisted electron transfer~\cite{gorman18,schlawin21}, spin-boson models~\cite{lemmer18}, and dynamics under vibronic-coupling Hamiltonians~\cite{macdonell21}.

Here, we describe a general approach to the quantum simulation of molecular spectra that avoids the exponential measurement requirements, instead requiring a number of measurements that is independent of molecular size. Our approach (Fig.~\ref{fig:map}e--f) uses analog quantum simulation techniques to predict the spectrum based on molecular dynamics in the time domain, unlike previous approaches that compute transition probabilities in the frequency domain (Fig.~\ref{fig:map}c--d).
We show how our scheme can be efficiently implemented on MQB simulators. The core of our method is a one-to-one mapping between molecular vibrational modes and bosonic degrees of freedom, and between molecular electronic states and qudit states, both available in several quantum-computing architectures.
Notably, the time-domain MQB approach allows for the inclusion of effects such as vibronic coupling, mixed states, and open quantum systems, all of which can have dramatic effects on the final spectrum, and are otherwise inaccessible in existing techniques.
We validate this approach by performing a proof-of-principle experimental demonstration by simulating a vibronic spectrum of \ce{SO2} using a trapped-ion quantum simulator, and achieving strong agreement between our experimental measurements, theory, and molecular spectroscopy measurements from the literature.

\begin{figure*}
    \centering
    \includegraphics[width=0.95\textwidth]{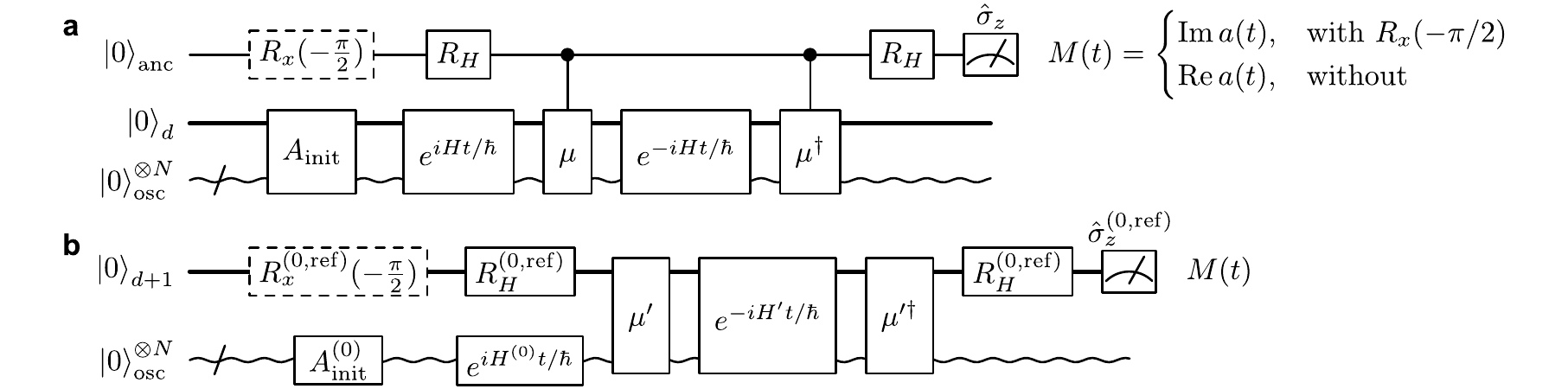}
    \caption{Quantum circuits for simulating vibronic spectra. 
    \textbf{a}, General circuit to obtain the autocorrelation function from an MQB device with $d$ qudit states (thick line) and $N$ bosonic modes (wavy line). By adding an ancilla qubit (top row) to the MQB simulator, the real and imaginary components of $a(t)$ can be measured as the $\hat{\sigma}_z$ expectation value of the ancilla, depending on whether the dashed-line gate $\hat{R}_x(-\pi/2)$ is applied or not. 
    $\hat{A}_\mathrm{init}$ prepares the initial state, and $\hat{R}_H$ is a Hadamard gate.
    \textbf{b}, Simplified circuit on an MQB simulator with an additional reference qudit state. $\hat{R}_H^{(0,\mathrm{ref})}$ and $\hat{R}_x^{(0,\mathrm{ref})}$ represent gates acting on qudit
    states $\ket{0}$ and $\ket{\mathrm{ref}}$, and the Hamiltonian $\hat{H}'$ and state preparation $\hat{\mu}'$ are modified to include $\ket{\mathrm{ref}}$. The initialisation $\hat{A}_\mathrm{init}^{(0)}$ and backwards time propagation $e^{i\hat{H}^{(0)}t/\hbar}$ act only on the initial electronic state. The expectation value of $\hat{\sigma}_z^{(0,\mathrm{ref})} = \ket{0}\bra{0} - \ket{\mathrm{ref}}\bra{\mathrm{ref}}$ leads to the same
    measurement outputs as in a.}
    \label{fig:circ}
\end{figure*}

\section{Time-domain spectroscopy on an analog quantum simulator}

Time-domain simulation avoids the individual measurement of the exponentially growing  number of spectroscopically relevant states~\cite{gordon65,cederbaum76,heller78,heller81}. The desired spectrum contains much less information than all the eigenstates of the molecule, and time-domain approaches can obtain it more directly. 

In the frequency-domain approach~\cite{huh15,huh17,shen18,sawaya19,wang20,jnane21}, the spectrum is a sum of individually calculated (or sampled) contributions from each eigenstate of the molecule, in proportion to how strongly they interact with light (Fig.~\ref{fig:map}c). For example, the first-order absorption spectrum of a molecule initially in eigenstate $\ket{\Psi_0}$ with frequency $\omega_0$ is \begin{equation} \label{eq:fspec}
    \sigma(\omega) = \sum_n \left|\braket{\Psi_n|\bm{\epsilon} \cdot \bm{\hat{\mu}}|\Psi_0}\right|^2 \delta(\omega - \omega_n + \omega_0),
\end{equation}
where $\omega$ is the frequency, $\ket{\Psi_n}$ is the eigenstate of the molecular Hamiltonian $\hat{H}$ with frequency $\omega_n$, $\bm{\hat{\mu}}$ is the dipole moment operator, and $\bm{\epsilon}$ is the polarisation of the light, both of which are vectors in three dimensions.
For simplicity, in what follows we write $\hat{\mu} = \bm{\epsilon} \cdot \bm{\hat{\mu}}$. 
The computational cost is exponential in the number of modes because of the need to calculate the exponentially many contributions in the sum, even if many peaks overlap or have zero intensity.
Although approaches that involve sampling of the wavefunction can reduce the number of measurements required~\cite{wang20}, the number of relevant eigenstates in the spectrum (i.e., the number of peaks) still grows exponentially with the number of modes. Therefore, obtaining an accurate spectrum requires an exponential number of samples to ensure that relevant features are not missed.

By contrast, the well-established time-domain view of spectroscopy (Fig.~\ref{fig:map}e) was developed to avoid having to calculate (originally on classical computers) all the eigenstates~\cite{gordon65,cederbaum76,heller78,heller81}. 
Eq.~\ref{eq:fspec} can be rewritten~\cite{heller81} using the Fourier definition of the delta function as
\begin{align} \label{eq:tspec}
    \sigma(\omega) &= \frac{1}{2\pi} \int_{-\infty}^{\infty}\mathrm{d}t\,e^{i\omega t}
    \braket{\Psi_0|\hat{\mu}^\dag e^{-i\hat{H}t/\hbar} \hat{\mu} e^{i\hat{H}t/\hbar}|\Psi_0} \nonumber\\
    &= \mathcal{F}\left[\braket{\Psi_\mu(0)|\Psi_\mu(t)}\right] = \mathcal{F}\left[a(t)\right],
\end{align}
where $\mathcal{F}$ is the Fourier transform, and $\ket{\Psi_\mu(t)} = e^{-i\hat{H}t/\hbar} \hat{\mu} e^{i\hat{H}t/\hbar} \ket{\Psi_0}$, i.e., the initial state evolved backwards in time for a time $t$, perturbed by the dipole operator, and evolved forwards in time $t$. 
The (dipole) autocorrelation function $a(t)$ is the overlap of the initial and time-evolved wavefunctions.
Eq.~\ref{eq:tspec} is a more general expression for the absorption spectrum than Eq.~\ref{eq:fspec}, since $a(t)$ is the dipole linear response function, $\braket{\mu^\dag(0)\mu(t)}$, which can also describe the evolution of mixed and non-stationary initial states, as well as open quantum systems.
It is a complex, Hermitian function, meaning that its Fourier transform is real and can be calculated with only $t>0$. 
Due to the Fourier relation between $a(t)$ and the spectrum, the spectral resolution of the simulation is determined by the maximum propagation time and the frequency range by the size of the time steps. Therefore, the cost of the simulation (i.e., the number of samples of $a(t)$ needed) is determined by the desired properties of the spectrum (resolution and range) and not by the number of eigenstates.

Therefore, $a(t)$ is the only quantity whose measurement is required to generate a vibronic spectrum.
It can be measured on a quantum simulator by keeping a copy of the initial wavefunction in superposition with the time-evolved wavefunction, so that their overlap can be determined at the end of the simulation via an interference measurement. 
Fig.~\ref{fig:circ}a shows the conceptually simplest approach to do so on a quantum simulator (either digital or analog) using the phase kickback technique with an ancilla qubit~\cite{abrams99,aspuruguzik05}, with a circuit that is similar to more general approaches to correlation-function measurement~\cite{terhal00,somma03,pedernales14}.
An operation $\hat{A}_\mathrm{init}$ prepares the initial state $\ket{\Psi_0}$ on the quantum simulator (or $\hat{\rho}_0$ for a mixed initial state), which
is then evolved by $e^{i\hat{H}t/\hbar}\ket{\Psi_0}$ (i.e., evolving with $-\hat{H}$ for a time $t$).
A Hadamard gate places the ancilla qubit in the superposition $\hat{R}_H \ket{0} = (\ket{0} + \ket{1})/\sqrt{2}$. The dipole operator is then controlled by the ancilla to give $(\ket{0}\otimes e^{i\hat{H}t/\hbar}\ket{\Psi_0} + \ket{1}\otimes\hat{\mu} e^{i\hat{H}t/\hbar}\ket{\Psi_0}) / \sqrt{2}$. Forward time evolution followed by a controlled $\hat{\mu}^\dag$ returns the $\ket{0}$ state to $\ket{\Psi_0}$, with a total state given by $(\ket{0}\otimes\ket{\Psi_0} + \ket{1}\otimes\hat{\mu}^\dag\ket{\Psi_\mu(t)}) / \sqrt{2}$.
The real part of $a(t)$ is then given by the expectation value of $\hat{\sigma}_z$ measurements on the ancilla, while the imaginary part can be obtained by inserting an additional $\hat{R}_x(-\pi/2) = e^{i\hat{\sigma}_x \pi/4}$ rotation before measurement, either before or after the two Hadamard gates. Equivalently, the ancilla can be used to control both of the time evolutions instead of the dipole operators.

The potentially impractical controlled unitary gates can, in most cases, be avoided if using MQB simulator.
We assume that the initial state $\ket{\Psi_0}$ is well described by a single electronic state, labelled $\ket{0}$, although extension to more electronic states is straightforward.
To measure $a(t)$ without an ancilla and controlled unitary gates, we add an additional electronic state to the simulator, i.e., for a simulation with $d$ electronic states, we require a $(d+1)$-level qudit for measurement of $a(t)$. We call this additional electronic state the reference state, $\ket{\mathrm{ref}}$, and use it to keep a copy of the initial wavefunction $\ket{\Psi_0}$ in superposition with the time-evolving wavefunction (Fig.~\ref{fig:circ}b).
This is achieved using the $\hat{R}_H^{(0,\mathrm{ref})}$ gate, which prepares the state 
$(\ket{0} + \ket{\mathrm{ref}})/\sqrt{2}$, where $\ket{0}$ is the electronic state of the initial wavefunction. $\ket{\Psi_0}$ is prepared on the bosonic modes by a single-electronic-state operation $\hat{A}_\mathrm{init}^{(0)}$, after which its initial evolution is simulated with $e^{i\hat{H}^{(0)}t/\hbar}$, where $\hat{H}^{(0)} = \bra{0} \hat{H} \ket{0}$ is the Hamiltonian describing evolution on only the initial electronic state.
The modified operator $\hat{\mu}'=\hat{\mu} + \ket{\mathrm{ref}}\bra{\mathrm{ref}}$ acts on the original $d$ qudit states, giving
$(\hat{\mu} e^{i\hat{H}^{(0)}t/\hbar}\ket{\Psi_0} + \ket{\mathrm{ref}}\otimes e^{i\hat{H}^{(0)}t/\hbar}\ket{\Psi_0})/\sqrt{2}$. This state then undergoes time evolution under the expanded Hamiltonian
\begin{equation}
    \hat{H}' = \hat{H} + \hat{H}^{(0)} \otimes \ket{\mathrm{ref}}\bra{\mathrm{ref}},
\end{equation}
so that the $\ket{\mathrm{ref}}$ component of the wavefunction returns to $\ket{\Psi_0}$ while the rest of the wavefunction propagates to $\ket{\Psi_\mu(t)}$. After the final $\hat{\mu}^\dag$, $a(t)$ is measured as the expectation value of $\hat{\sigma}_z^{(0,\mathrm{ref})} = \ket{0}\bra{0} - \ket{\mathrm{ref}}\bra{\mathrm{ref}}$ (with the $\hat{R}_x^{(0,\mathrm{ref})}$ gate differentiating between the real and imaginary parts).

In either scheme, the simulation needs to be repeated sufficiently many times to numerically converge both the real and imaginary parts of $a(t)$ to the required precision for a discrete number of times $t$.
It is possible to halve the number of measurements because the hermiticity of $a(t)$ implies that the spectrum can be reconstructed from only $\mathrm{Re}\,a(t)$ (see Appendix~\ref{app:posw}).

An MQB device such as a trapped-ion or cQED system can simulate a wide range of realistic molecular Hamiltonians~\cite{macdonell21}.
Specifically, an MQB simulator with second-order light-matter interactions~\cite{katz22} can simulate a quadratic vibronic-coupling (QVC) Hamiltonian~\cite{macdonell21},
\begin{align}
    \hat{H}_\mathrm{QVC} &= \hat{H}_0 + \sum_{n,m=0}^{d-1} \hat{C}_{n,m} \ket{n}\bra{m}, \label{eq:qvc}
\end{align}
which includes $N$ free harmonic oscillators, $\hat{H}_0 = \sum_{j=1}^N \hbar\omega_j (\hat{n}_j + \tfrac{1}{2})$, and expansion coefficients
\begin{align}
    \hat{C}_{n,m} &= c_0^{(n,m)} + \sum_{j=1}^N c_j^{(n,m)} \hat{Q}_j + \sum_{j,k=1}^N c_{j,k}^{(n,m)} \hat{Q}_j\hat{Q}_k,
\end{align}
describing perturbations of individual electronic potential energy surfaces ($n = m$) and vibronic couplings between them ($n \neq m$). In these equations, $\hat{Q}_j = (M_j \omega_j/\hbar)^{1/2} \hat{q}_j = (\hat{a}_j^\dag + \hat{a}_j)/\sqrt{2}$ are the nuclear normal modes $\hat{q}_j$ weighted by reduced mass $M_j$ and frequency $\omega_j$,
$\hat{n}_j = \hat{a}_j^\dag \hat{a}_j$ are the number operators, $\hat{a}_j^\dag$ and $\hat{a}_j$ are the bosonic creation and annihilation operators, $\ket{n}$ are the electronic states, and $c$ are the vibronic expansion terms. 

For the simulation of a QVC model spectrum, $\hat{H}^{(0)}$ can usually be chosen to be equal to $\hat{H}_0$. Mathematically, there are two requirements for this to be met. First, as is usually the case in molecular systems, the ground electronic state should have negligible coupling to excited ones, $\hat{C}_{n,0}=\hat{C}_{0,n}=0$. Second, without loss of generality, we can choose $\hat{C}_{n,m}$ such that $\hat{C}_{0,0}=0$.

\section{Example: One-mode model of \ce{SO2}}

\begin{figure}
    \centering
    \includegraphics[width=0.95\columnwidth]{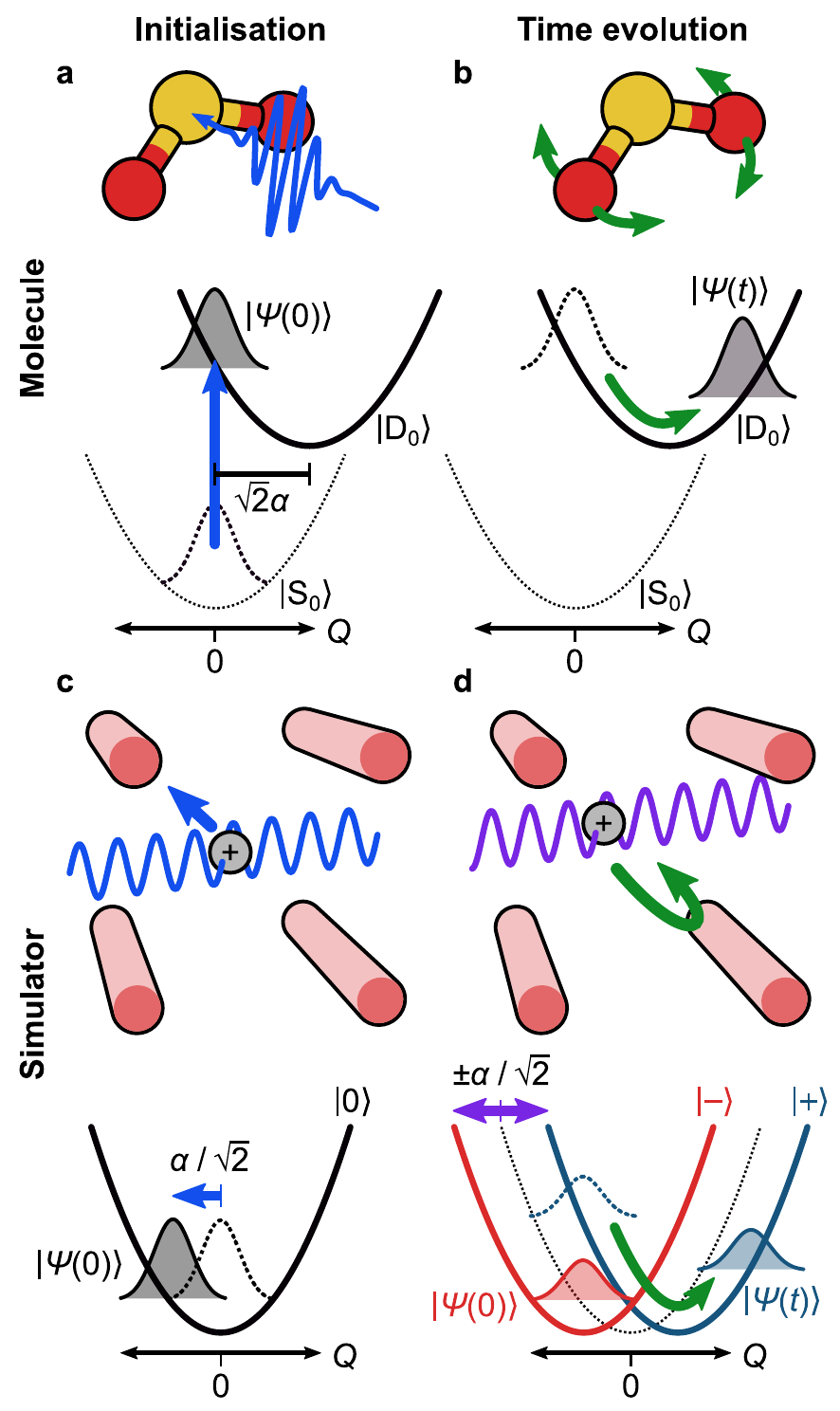}
    \caption{The trapped-ion simulation of the \ce{SO2} photoelectron spectrum. 
    \textbf{a}, Photoexcitation of \ce{SO2} from the neutral ground electronic state (S$_0$)
    leads to the ground cationic electronic state (D$_0$), at a geometry that is $\sqrt{2}\alpha$ from the D$_0$ minimum along the vibrational mode. 
    \textbf{b},~Time evolution on D$_0$ (green arrow) corresponds to bending of the \ce{SO2} molecule. 
    \textbf{c}, In an ion-trap MQB simulator, the initial state is prepared by displacing the ion position by $-\alpha/\sqrt{2}$ with a laser-ion interaction (blue arrow).
    \textbf{d}, Time evolution of the simulator (green arrow) happens under a laser-induced spin-motion interaction that causes the effective potentials to be shifted by $\pm\alpha/\sqrt{2}$ in the $\ket{\pm}$
    basis (purple arrows).}
    \label{fig:so2_concept}
\end{figure}

The simplest example of our approach involves dynamics on a single electronic state with a single vibrational mode.
This type of model can be used to describe the photoelectron spectrum of the $\mathrm{S}_0 \rightarrow \mathrm{D}_0$ transition of \ce{SO2} (Fig.~\ref{fig:so2_concept}a),
which has a bending mode with frequency $\omega_\mathrm{b} = 2\pi \times \SI{12.44}{THz}$, along which the D$_0$ electronic potential energy surface
has a displacement relative to S$_0$ given by $\alpha = 1.716$~\cite{lee09} (in unitless, mass- and frequency-scaled coordinates).
This model is described by the Hamiltonian
\begin{multline}
    \label{eq:minimal}
    \hat{H}_{\ce{SO2}} = \hbar\omega_\mathrm{b}\hat{n} 
    + E_{\mathrm{S}_0}\ket{\mathrm{S}_0}\bra{\mathrm{S}_0}\\
    + (E_{\mathrm{D}_0} - \sqrt{2}\hbar \omega_\mathrm{b} \alpha \hat{Q})\ket{\mathrm{D}_0}\bra{\mathrm{D}_0},
\end{multline}
where $E_{\mathrm{S}_0}$ and $E_{\mathrm{D}_0}$ are the potential energies of S$_0$ and D$_0$ at $Q = 0$, and we removed the (constant) zero-point energy $\hbar\omega_\mathrm{b}/2$.
For approximations involved, see Sec.~\ref{sec:methods}.

Under the Condon approximation, we assume that the electronic (dipole) transition completely transfers the population from S$_0$ to D$_0$ with no effect on the nuclear coordinates, i.e., $\hat{\mu} = \ket{\mathrm{D}_0}\bra{\mathrm{S}_0} + \text{h.c.}$. After transitioning to D$_0$, the wavefunction is no longer stationary and the molecule begins to vibrate (Fig.~\ref{fig:so2_concept}b).

To measure the autocorrelation function of \ce{SO2} on an MQB simulator, we add a reference state to the model, with corresponding Hamiltonian $\hat{H}^{(0)}\otimes\ket{\mathrm{ref}}\bra{\mathrm{ref}}$, where $\hat{H}^{(0)} = \hat{H}_0 = \hbar\omega_\mathrm{b}\hat{n}$. This results in a Hamiltonian with three electronic states: $\ket{\mathrm{S}_0}$, $\ket{\mathrm{D}_0}$, and $\ket{\mathrm{ref}}$. However, because $\hat{\mu}$ causes 100\% population transfer to D$_0$ and $\hat{H}_{\ce{SO2}}$ contains no terms that couple the two electronic states, we can remove S$_0$ from the model completely, replacing $E_{\mathrm{D}_0}$ with $\Delta E = E_{\mathrm{D}_0} - E_{\mathrm{S}_0}$ to conserve the excitation energy. We can also remove the initial time evolution $e^{i\hat{H}^{(0)}t/\hbar}$, since $\ket{\Psi_0}$ is a stationary state of $\hat{H}^{(0)}$.
The two remaining electronic states can be represented by a qubit, $\ket{0}=\ket{\mathrm{D}_0}$ and $\ket{1}=\ket{\mathrm{ref}}$, giving
\begin{multline}
    \hat{H}'_{\ce{SO2}} = \big(\hbar\omega_\mathrm{b}\hat{n} - \sqrt{2}\hbar\omega_\mathrm{b}\alpha\hat{Q} + \Delta E\big) \ket{0}\bra{0} \\
    + \hbar\omega_\mathrm{b}\hat{n} \ket{\mathrm{1}}\bra{\mathrm{1}}, \label{eq:hpso2}
\end{multline}
which corresponds to Eq.~\ref{eq:qvc} with $\omega_1 = \omega_\mathrm{b}$, $c_1^{(0,0)} = \sqrt{2} \hbar \omega_b \alpha$, and all other coefficients equal to zero.
In this representation, the corresponding dipole operator is $\hat{\mu}' = \mathbb{1}$, because the $\ket{\mathrm{D}_0}$ electronic state corresponds to the initial qubit state $\ket{0}$.

For experimental implementation, $\hat{H}'_{\ce{SO2}}$ can be transformed into a more convenient, but equivalent, form (even though Eq.~\ref{eq:hpso2} is already in the general form suitable for MQB simulation). First, we remove the constant term $\Delta E$ on $\ket{0}$, which is the initial excitation energy of the wavefunction from S$_0$ to D$_0$; doing so leads to a constant frequency shift of the entire spectrum, which can be restored by adding $\Delta E/\hbar$ to the frequencies after the spectrum is predicted.
Next, we transform the Hamiltonian into a form that is symmetric about $Q = 0$. In $\hat{H}'_{\ce{SO2}}$, the minimum of the D$_0$ potential energy surface is at $Q = \sqrt{2}\alpha$ and that of the reference state is at $Q = 0$. These minima can be made symmetric in a displaced coordinate system obtained using the displacement operator $\hat{D}(-\alpha/2) = e^{-\alpha(\hat{a}^\dag + \hat{a})/2}$. Finally, the two Hadamard gates can be incorporated into the time evolution. Altogether, this gives
\begin{align} \label{eq:so2pp}
    \hat{H}_{\ce{SO2}}'' &=
    \hat{R}_H \hat{D}(-\alpha/2) \hat{H}'_{\ce{SO2}} \hat{D}^\dag(-\alpha/2) \hat{R}_H \nonumber\\
    &= \hbar\omega_\mathrm{b} \hat{n} + \frac{\hbar\omega_\mathrm{b}\alpha}{\sqrt{2}} \hat{\sigma}_x \hat{Q}.
\end{align}
which is a Jaynes-Cummings-type interaction that we implement experimentally below.
As a result of the Hadamard transformation, $\ket{\mathrm{D}_0}$ and $\ket{\mathrm{ref}}$ are now $\ket{+}$ and $\ket{-}$ (Fig.~\ref{fig:so2_concept}d), where $\ket{\pm} = (\ket{0} \pm \ket{1})/\sqrt{2}$.

\begin{figure*}
    \centering
    \includegraphics[width=0.95\textwidth]{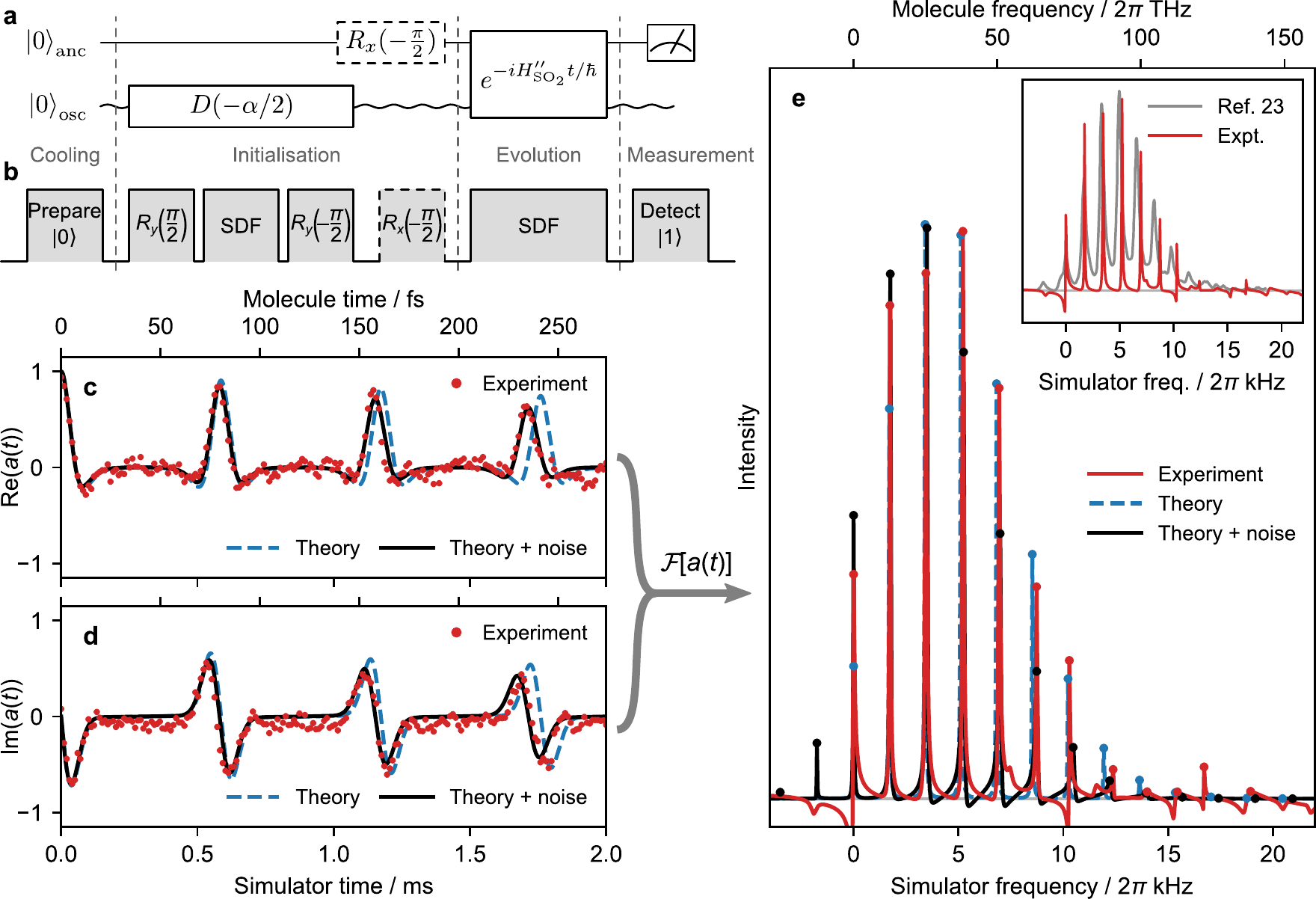}
    \caption{Experimental time-domain simulation of the single-mode \ce{SO2} vibronic spectrum. 
    \textbf{a}, Quantum circuit diagram for the simulation to extract the real and imaginary components of $a(t)$, using one qubit and one bosonic mode. \textbf{b}, Experimental pulse sequence implementing the quantum circuit. 
    \textbf{c} and \textbf{d}, Simulations and measurements of $a(t)$. ``Theory\,+\,noise'' indicates a simulation accounting for known sources of noise.
    \textbf{e}, Comparison of the Fourier transformed data from c and d with theoretical predictions. Dots indicate peak maxima. Inset: comparison of the spectroscopically observed spectrum at \SI{320}{K}~\cite{holland94} with frequencies shifted to give $E_{\mathrm{D}_0} = 0$ and the ion-trap experiment.
    The decreasing peak spacing in the \ce{SO2} spectrum is caused by a weak anharmonicity that is neglected in the single-mode model.}
    \label{fig:so2_Data}
\end{figure*}

The overall circuit for this simulation is shown in Fig.~\ref{fig:so2_Data}a. 
The initialisation consists of a $\hat{A}_\mathrm{init}^{(0)} = \hat{D}(-\alpha/2)$ operator, which displaces the initial vibrational ground state into the same displaced coordinates as the Hamiltonian. 
The time evolution consists of the unitary $e^{-i\hat{H}''_{\ce{SO2}}t/\hbar}$. 
The measurement of the real part of $a(t)$ proceeds directly from the qubit state, using the operator $\hat{\sigma}_z^{(0,\mathrm{ref})} = \hat{\sigma}_z$. As before, the imaginary part requires the additional $\hat{R}_x(-\pi/2)$ gate.

\section{Experimental quantum simulation}

We experimentally demonstrate the one-mode \ce{SO2} simulation using a trapped-ion MQB quantum simulator~\cite{Milne:2022}. Our system confines a single \ce{^171Yb+} ion in a linear Paul trap, and we encode a qubit in the ion's $\LS{S}{1/2}$ hyperfine ground-state manifold, $\ket{0}\equiv\ket{F=0,m_F=0}$ and $\ket{1}\equiv\ket{F=1,m_F=0}$. The ion's vibration in the transverse $x$ direction encodes the molecular vibration.

The key tool for manipulating the qubit and motional wavepacket of the ion is a pair of Raman laser beams. As detailed in Sec.~\ref{sec:methods}, a bichromatic laser pulse can apply a state-dependent displacement force (SDF) to the ion, described in the interaction picture by the Hamiltonian
\begin{equation}
    \hat{H}_{\mathrm{SDF}}^I = \hbar \Omega_\mathrm{S} \hat{\sigma}_x (\hat{a}^\dag e^{i(\delta t + \varphi)} + \mathrm{h.c.}),
\label{eq:MSgate}
\end{equation}
where the three adjustable parameters are the motional sideband interaction strength $\Omega_\mathrm{S}$, the detuning of the bichromatic components $\delta$, and the motional phase $\varphi$. In the Schr{\"o}dinger picture, this Hamiltonian takes the time-independent form
\begin{equation}
    \hat{H}_{\mathrm{SDF}} = \hbar\delta\hat{n} + \sqrt{2}\hbar\Omega_\mathrm{S} \hat{\sigma}_x\big(\cos(\varphi) \hat{Q} + \sin(\varphi) \hat{P}\big),
\end{equation}
where $\hat{P}$ is the conjugate momentum of $\hat{Q}$. This equation is identical to Eq.~\ref{eq:so2pp} when $\varphi = 0$, $\delta = \omega_\mathrm{b}$ and $\Omega_\mathrm{S} = \omega_\mathrm{b}\alpha/2$. In practice, these values need to be scaled from molecular frequencies to experimental frequencies by a constant scaling factor $F$ whose value depends on the type of MQB simulator~\cite{macdonell21}.

The experimental pulse sequence, Fig.~\ref{fig:so2_Data}b, describes the four stages of the simulation protocol: (i) cooling, (ii) initialisation, (iii) time evolution, and (iv) measurement.

Cooling prepares the ion in the ground state, from which further operations can be executed. First, Doppler cooling and sideband cooling are used to cool the motional degree of freedom as close as possible to the ground state (we obtained $\bar{n}\approx0.05$). Second, optical pumping on the internal electronic state is used to prepare the qubit state $\ket{0}$. For details, see Sec.~\ref{sec:methods}.

Initialising the \ce{SO2} simulation requires preparing the state $\ket{0} \otimes \ket{-\alpha/2}$, where the first ket refers to the qubit and the second to the displaced motional ground state. In the three-pulse initialisation sequence, the first $\hat{R}_y(\pi/2)$ pulse rotates the qubit to the $\ket{+}$ state while the second $\hat{R}_y(-\pi/2)$ pulse returns the qubit to the $\ket{0}$ state. Between the two rotations, $\hat{H}_{\mathrm{SDF}}$ is applied on resonance ($\delta=0$, $\varphi = -\pi/2$) for \SI{0.093}{ms}, implementing the operation $\ket{+} \otimes \ket{0}$ $\rightarrow$ $\ket{+}\otimes \ket{-\alpha/2}$ with $\alpha/2 = 0.858$. The overall sequence produces the desired $\ket{0} \otimes \ket{-\alpha/2}$. For details, see Sec.~\ref{sec:methods}.

Time evolution is accomplished using $\hat{H}_{\mathrm{SDF}} = F \hat{H}''_{\ce{SO2}}$ with $\varphi = 0$. We use the scaling factor $F = 1.37 \times 10^{-10}$ to convert molecular timescales and frequencies (fs, THz) to trapped-ion timescales and frequencies (ms, kHz).
Doing so gives $\delta = F\omega_\mathrm{b} = 2\pi \times \SI{1.705}{kHz}$ and $\Omega_\mathrm{S} = F\omega_\mathrm{b}\alpha/2 = 2\pi \times \SI{1.463}{kHz}$.
To obtain the time trace of $a(t)$, we measure its value at 200 different times $t$ by repeating the experiment with the duration of unitary evolution under $\hat{H}_{\mathrm{SDF}}$ varying between 0 and \SI{2}{ms}, corresponding to molecular durations of up to \SI{274}{fs}.

The final step in the simulation is the measurement of $a(t)$, which is carried out by measuring the qubit in the computational basis (for details, see Sec.~\ref{sec:methods}).
Reading out the imaginary part of $a(t)$ requires the additional $\hat{R}_x(-\pi/2)$ gate on the qubit following the displacement in the initialisation step.

The full experimental sequence above is repeated 500 times for each duration $t$ of the simulated time evolution in order to converge the measurement observables. 

Fig.~\ref{fig:so2_Data}c--e shows the agreement between our experimental results and theoretical predictions. The predicted and measured $a(t)$ are shown in Fig.~\ref{fig:so2_Data}c--d. The theoretical curve is calculated as shown by the circuit in Fig.~\ref{fig:so2_Data}a, using Eq.~\ref{eq:so2pp} for the time evolution. 
To give a non-zero linewidth in the theoretical spectrum, the predicted $a(t)$ was multiplied by an exponential decay of \SI{6}{ms} (corresponding to \SI{822}{fs} at the molecular timescale). 
Fig.~\ref{fig:so2_Data}e shows the agreement between the predicted and measured spectra, i.e., the Fourier transforms of the theoretical and experimental $a(t)$.

Despite the good overall agreement between theory and experiment, there are minor differences between the two, most of which can be explained by the presence of noise in the trapped-ion simulator.
Points of difference include a drift in the $a(t)$ signal with simulation time, discrepancies in peak heights, and asymmetric lineshapes.
Most of the discrepancies can be accounted for by adding a model of experimental noise to our theory. This simulation includes a linear frequency drift with the Hamiltonian $\hat{H}_{\ce{SO2}}^{\mathrm{fit}} = \hat{H}''_{\ce{SO2}} + d_\delta \hat{n} t$ and uses an initial thermal state with an average motional state population $\bar{n}$. The evolution of the density operator $\hat{\rho}$ obeys the master equation
\begin{equation}
    \frac{\partial\hat{\rho}}{\partial t} = -\frac{i}{\hbar} [\hat{H}_{\ce{SO2}}^{\mathrm{fit}}, \hat{\rho}] + \Big(\gamma_h \mathcal{D}[\hat{a}^\dag] + \frac{2}{\tau_d} \mathcal{D}[\hat{n}]\Big) \hat{\rho},
\end{equation}
where $\mathcal{D}[\hat{L}]\hat{\rho} = \hat{L}\hat{\rho}\hat{L}^\dag - \frac{1}{2} \{\hat{L}^\dag\hat{L},\hat{\rho}\}$ is a Lindblad superoperator acting on $\hat{\rho}$ for the jump operator $\hat{L}$. The dissipation is described by a motional heating rate $\gamma_h$ and a pure motional dephasing lifetime $\tau_d$. 
The noise parameters are fitted to the measured $a(t)$ using non-linear least squares, giving $d_\delta = 2\pi \times \SI{52}{Hz~ms^{-1}}$, $\bar{n} = 0.061$, $\gamma_h = \SI{43}{s^{-1}}$ and $\tau_d = \SI{110}{ms}$. This yields an effective motional coherence time of \SI{33}{ms}, in agreement with the experimentally measured value (see Sec.~\ref{sec:methods}). The simulation that includes noise agrees better with the experiment, accounting for the signal drift in $a(t)$ using the frequency drift (Fig.~\ref{fig:so2_Data}c--d), as well as linewidths and peak heights using $\bar{n},$ $\gamma_h$ and $\tau_d$ (Fig.~\ref{fig:so2_Data}e).

\section{Discussion}

Our approach has two types of advantages over existing proposals for the quantum simulation of spectroscopy: those which result from our theoretical framing of the simulation, and those that come from our choice of experimental platform.

The advantages of our theoretical framework stem from framing molecular spectroscopy in the time domain. Our work mirrors the development of classical computing methods in the time domain, which greatly simplified the calculation of spectra for high-dimensional systems without the need to resolve eigenvalues~\cite{heller78,heller81}. In the time domain, spectroscopy is an initial-value problem, rather than an eigenvalue problem where the number of solutions grows exponentially with system size. This reframing leads to two distinct advantages over competing proposals for the analog quantum simulation of spectroscopy: scalability and generality. 

The scalability of our approach stems from the exponentially reduced number of measurements needed to predict the spectrum. In frequency-domain approaches, the number of eigenvalues (i.e., peaks) grows exponentially with the number of modes (i.e., with molecule size), each of which needs to be sampled to determine its intensity. Even if the number of eigenvalues is truncated on an ad hoc basis, the number of significant eigenvalues grows rapidly. For example, the technique employed by Shen et al.~\cite{shen18} involves a sequence of laser pulses to project the population of each multimode motional state $\ket{n_1, n_2, ..., n_N}$ onto the qubit state population; if each mode occupation is truncated at $n_\mathrm{max}$, the computational cost scales exponentially as $n_\mathrm{max}^N$. By contrast, in time-domain approaches such as ours, the number of measurements required is independent of system size. Instead, the number of measurements is determined by the desired frequency range and resolution of the spectrum, which are the inverses of the time step and the total simulation time, respectively. Perfect spectral resolution is not necessary for characterising a spectrum, since measured spectra of even modestly sized molecules have broad features of overlapping peaks, especially when environmental effects, strong coupling, and limited measurement resolution are considered. Therefore, the cost of a time-domain simulation is determined by experimentally relevant parameters (spectral range and resolution), rather than the size of the underlying Hamiltonian.

As for generality, our method can be used to predict the spectroscopy of any chemical system due to the fully general relationship between $\sigma(\omega)$ and $\mathcal{F}[a(t)]$ shown in Eq.~\ref{eq:tspec}. The observable we measure, $a(t) = \braket{\mu^\dag(0)\mu(t)}$~\cite{cederbaum76,heller81}, is defined without an eigenstate expansion and can, in principle, be efficiently measured on any quantum simulator, including those simulating open quantum systems, vibronic couplings, nonlinearities, or non-Condon effects.

The simulation of open quantum system is the most striking example of the generality of the time-domain approach. Introducing controlled noise into a simulation allows an MQB simulator to simulate environmental effects such as peak broadening~\cite{macdonell21} with the same number of measurements  of $a(t)$. By contrast, frequency-domain approaches typically fail on open systems, which no longer have discrete eigenstates that can be measured one by one. Furthermore, in the time-domain approach, initial conditions can likewise include mixed states such as thermal states, allowing simulations of spectroscopy of molecules at finite temperature without additional experimental resources. By contrast, doing so in the frequency domain requires multiple experiments~\cite{huh15} or doubling the simulator size~\cite{huh17}.

The generality of our approach also extends to the ability to include vibronic couplings (Eq.~\ref{eq:qvc}, $n \neq m$) and non-Condon effects. Vibronic coupling is ubiquitous in UV-visible spectroscopy, and, in the time domain, any approach able to simulate dynamics with vibronic coupling~\cite{macdonell21} can also predict the spectrum. In contrast, all frequency-domain analog simulation approaches use Duschinsky transformations to prepare the initial state in the vibrational basis of the final electronic state~\cite{huh15,huh17,shen18,wang20,jnane21}. This is a powerful technique for energetically separated electronic states, but cannot describe vibronic coupling. 
In addition, our approach has the potential to simulate non-Condon effects (the dependence of $\hat{\mu}$ on nuclear coordinates) with no additional experimental resources. In comparison, non-Condon effects require multiple simulations for current frequency-domain approaches~\cite{jnane21}.

Turning to the advantages of our experimental demonstration, the use of MQB simulators offers a significant reduction in required quantum resources over digital quantum simulation approaches. Both analog MQB and digital simulations of dynamics---either of which could be used for our time-domain simulation of spectroscopy---require resources that scale linearly with system size~\cite{macdonell21}; however, the resource cost per mode is considerably higher in digital approaches, where many qubits would be needed to adequately represent a single vibrational mode. For example, our demonstration of \ce{SO2} with a single trapped ion is equivalent to a digital encoding with at least 6 qubits (assuming 32 Fock states per mode). MQB simulation also comes with a time advantage, since the harmonic motion native to an MQB simulator would require many gates to implement digitally. Finally, the relative frequencies of qudit and bosonic levels on a trapped-ion MQB simulator lead to relative  electronic and vibrational noise strengths that are similar to those in molecules, which can be exploited for simulating open quantum systems and which would not be the case for digital simulation, where all qubits are ordinarily assumed to experience comparable noise.

Our experimental results demonstrate the availability of the essential MQB building blocks in existing trapped-ion technology. Our demonstration is a proof of principle, and further work is necessary to reach a scale where it could outperform classical computers by simulating larger, more complicated molecules. Indeed, Eq.~\ref{eq:minimal} captures the absorption of a single displaced harmonic oscillator, and can be solved analytically~\cite{cederbaum76}. 
Nevertheless, we see a clear path towards integrating complicated, non-linear vibronic-coupling Hamiltonians into analog spectroscopy simulations of molecules large enough to be intractable on classical computers.
All of the necessary components for a more general, QVC simulation~\cite{macdonell21} have already been demonstrated in trapped-ion systems, including those with qudits and more vibrational modes. 
In particular, our approach can incorporate more modes using an additional Raman interaction for every mode, which can be efficiently implemented with the same experimental setup by interleaving different interactions using Trotterisation~\cite{lloyd95,macdonell21}, an established technique in trapped ions~\cite{lanyon_universal_2011}.  
Furthermore, higher-order terms in the vibronic-coupling Hamiltonian---responsible for anharmonicities and nonlinearities that are particularly difficult to simulate classically---can be incorporated into both the simulated Hamiltonian and the initial state preparation using techniques such as motional squeezing and mode-mixing~\cite{shen18,marshall16,katz22} or using ancillary ions~\cite{gerritsma11}.
In addition, more electronic states could be simulated using recent experimental advances in trapped-ion qudits~\cite{Low2020,ringbauer_universal_2022}. 

As in any analog simulation---quantum or classical---the absence of error correction means that excessive noise can lead to inadequate results. However, our demonstration shows that, despite the lack of error correction, existing trapped-ion technology can provide remarkable agreement with theoretical predictions.
Moreover, our analysis of experimental noise sources shows that most of the imperfections in our simulation can be accounted for, making it clear which experimental improvements are necessary if greater accuracy is desired.
When simulating larger molecules or those open to the environment---where classical chemical simulations struggle the most---the presence of noise in the simulator becomes a feature that can be controlled. For instance, the inset in Fig.~\ref{fig:so2_Data}e shows that our simulation gives narrower peaks than are measured in a high-precision spectroscopic experiment at \SI{320}{K}, meaning that we would have to inject additional noise to fully reproduce the spectroscopic observations.

Our method's most likely path to quantum advantage is by simulating a combination of effects that make classical simulation challenging, including vibronic coupling and a finite-temperature bath. We previously outlined~\cite{macdonell21} the favourable resource scaling that our approach can achieve for such systems. For example, a full-dimensional quadratic vibronic coupling model of pyrazine is a challenging system for classical computers~\cite{raab99}. It involves 24 modes and two electronic states, meaning that our technique could simulate its spectrum with 8 trapped ions~\cite{macdonell21}. For comparison, previous ion-trap experiments have controlled interactions of as many as 20 ions~\cite{friis18}, putting our example within existing experimental feasibility.

Overall, our approach shows the remarkable advantages---both theoretical and experimental---of using the time-domain representation of spectroscopy in analog quantum simulation. By using a one-to-one mapping between simulated molecular vibrations and bosonic modes in a quantum simulator, our scheme provides an exponential improvement in resource requirements compared to existing quantum methods using frequency-domain simulations. In addition, it straightforwardly generalises to simulations of more complicated chemical systems or those open to the environment. Our proof-of-principle demonstration of the simplest example of our approach showcases all of the necessary experimental building blocks, giving us confidence that foreseeable developments in quantum technology will allow larger simulations of molecular spectroscopy to occur in the near term, including of molecules that could not be simulated on any classical computer.

\section{Methods}
\label{sec:methods}

\paragraph*{Hamiltonian for \ce{SO2}.}
We selected \ce{SO2} for our proof-of-principle demonstration because it is well described by the  Hamiltonian in Eq.~\ref{eq:minimal}.
This Hamiltonian includes several simplifications to the physics of \ce{SO2}, all of which could be relaxed in a more detailed model. First, the model assumes equal frequency $\omega_\mathrm{b} = 2\pi \times \SI{12.44}{THz}$ for both electronic states, although the ground electronic state has a slightly higher vibrational frequency of $2\pi \times \SI{15.55}{THz}$~\cite{lee09}.
Second, it neglects higher-lying electronic states (and vibronic couplings to them) due to their large energetic separation from S$_0$ and D$_0$.
Finally, only the bending mode is included because the displacements between S$_0$ and D$_0$ are small along the other two vibrational modes, the symmetric and asymmetric stretches, being $\alpha = -0.026$ and $\alpha = 0$, respectively~\cite{lee09}.

\paragraph*{Numerical methods.}
Time evolution for all theoretical simulations was simulated using the master equation solver in QuTiP~\cite{qutip}. Curve fitting was performed with the Levenberg-Marquardt algorithm in SciPy~\cite{scipy}. All Fourier transforms were calculated using the Fourier-Pad\'{e} approximation (FPA). Unlike a discrete Fourier transform (DFT), the FPA finds coefficients of continuous rational functions that approximate the Fourier transform from discrete data. Relative to DFT, it has a faster convergence for features of a spectrum generated from a finite-length time series~\cite{bruner16}. To avoid poles due to the rational function expansion, $a(t)$ was multiplied by a frequency shift $e^{i\theta t}$ with $\theta = 2\pi \times \SI{8}{kHz}$ before the Fourier transform, which was later corrected by shifting the spectrum by $-\theta$. Testing values of $\theta$ in the range $2\pi \times 7\text{--}\SI{9}{kHz}$ showed no change in the spectrum, indicating no poles or spurious peaks.

\paragraph*{Ion trap characteristics.} 
The motional mode frequencies of our ion-trap MQB simulator are $\{\omega_x,\omega_y,\omega_z\} = 2\pi \times \{1.31, 1.45, 0.5\}$ MHz. Using Ramsey-type measurements, we find native (uncorrected) coherence times of $T_2^* = \SI{8.7}{s}$ for the qubit and an effective $\SI{35}{ms}$ for the transverse motional mode along $x$. We measured a motional mode rethermalisation (heating) rate of \SI{0.2}{quanta~s^{-1}} in the absence of laser light. 

\paragraph*{Coherent operations.} 
The qubit states and motional modes are manipulated by stimulated Raman transitions driven by a \SI{355}{nm} pulsed laser~\cite{optical_comb,Islam:14}. Two separately controllable \SI{355}{nm} laser beams are focused on the ion's location; they form an orthogonal geometry such that only the $x$ and $y$ transverse modes of motion can be driven. The applied laser light is controlled using acousto-optic modulators (AOM), driven by radio-frequency (RF) signals. Changing the RF signal amplitude, frequency and phase allows the tuning of $\Omega_\mathrm{S}$, $\delta$ and $\varphi$ in Eq.~\ref{eq:MSgate}, respectively. These parameters are controlled using RF signal generators as part of the experiment control system~\cite{ARTIQ:2021}.

\paragraph*{Cooling and qubit state preparation.} 
A laser, red-detuned from the $\LS{S}{1/2} \rightarrow \LS{P}{1/2}$ transition near \SI{369.5}{nm}, is used to Doppler cool the motional modes to a thermal state. The Doppler-cooled ion temperature is further reduced with pulsed resolved-sideband cooling \cite{Diedrich.1989,sideband_cooling}. 
The qubit state is prepared by optically pumping to $\ket{0}$ with a beam resonant with the $\LS{S}{1/2} \rightarrow \LS{P}{1/2}$ transition.
The $\LS{P}{1/2}$ state has a non-zero probability of decaying which results in population in $\LS{D}{3/2}$ and $\LS{F}{7/2}$ states; this leaves the ion state off-resonant from the Doppler cooling laser. Therefore, additional light fields at \SI{935}{nm} and \SI{760}{nm} are used to depopulate these states and return the ion to the cooling cycle~\cite{Olmschenk:2007,Edmunds:2021}.

\paragraph*{SDF interaction.} 
Using a two-tone RF signal to drive one of the \SI{355}{nm} laser beams, a bichromatic light field is generated, which can simultaneously drive the red- and blue motional sideband transitions adjacent to the qubit resonance. This bichromatic light creates the state-dependent force described by $\hat{H}_\mathrm{SDF}$, which acts in the $\hat{\sigma}_x$ eigenbasis of the qubit~\cite{molmer99}. Through a prior rotation of the internal qubit state into the eigenbasis, the SDF enacts a coherent displacement of the motional wavepacket in a particular direction in phase space~\cite{Milne:2021}. For details of the calibration procedure, see Appendix~\ref{app:calibration}.

\paragraph*{Measurement.} The qubit state is measured using state-dependent fluorescence, induced by a laser beam resonant with the $\LS{S}{1/2}\ket{1} \rightarrow \LS{P}{1/2}\ket{F=0}$ transition~\cite{Olmschenk:2007,Blinov:2004}. The $\ket{0}$ state is off-resonant from the detection beam by roughly $2\pi\times$\SI{14.75}{GHz} and only scatters an average of 0.1 photon counts during a detection window. Measurement of the $\ket{1}$ state produces 16.7 photon counts. Using a threshold of 4 counts we are able to infer the qubit state with a detection error of 1.2\%~\cite{Edmunds:2021}.

\section*{Acknowledgements}
We were supported by a Westpac Scholars Trust Research Fellowship (I.K.), by the Lockheed Martin Corporation, by the Australian Government's Defence Science and Technology Group, by the United States Office of Naval Research Global (N62909-20-1-2047), by the Sydney Quantum Academy (T.R.T.), by the University of Sydney Nano Institute, and by the Australian National Computational Infrastructure.

\section*{Author contributions}
RJM designed the quantum algorithm and performed theoretical simulations. TN performed the experiment. TN, TFWR, CHV, ADR, MJM, and TRT developed the experimental system. MAC simulated experimental noise. MJB, TRT, and CH supervised the experimental work. IK supervised the theoretical work. RJM and TN drafted the manuscript. All authors refined the project design and edited the manuscript.

\bibliography{vc_spectra.bib}

\section*{APPENDICES}

\appendix

\section{Simulation of positive-frequency spectra}
\label{app:posw}
The most general approach to measure $a(t)$ with an MQB simulator requires two experiments at each time $t$: one to measure $\mathrm{Re}\,a(t)$, and another to measure $\mathrm{Im}\,a(t)$. However, the total number of measurements can be halved by exploiting the hermiticity of $a(t)$ to only measure $\mathrm{Re}\,a(t)$.

\begin{figure}
    \centering
    \includegraphics[width=\columnwidth]{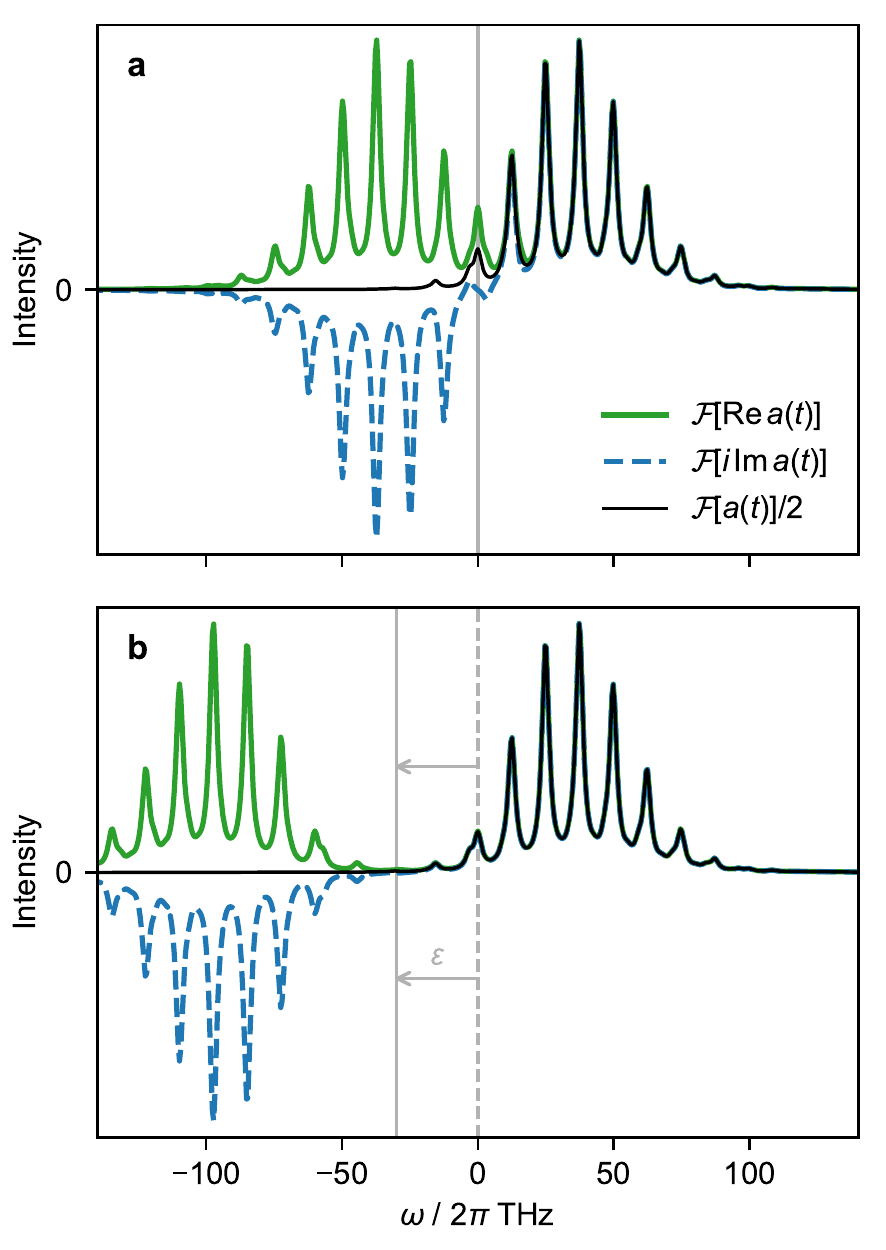}
    \caption{Fourier transforms of different components of the autocorrelation function $a(t)$, using the D$_0$ photoelectron spectrum of \ce{SO2} as an
    example~\cite{lee09}. \textbf{a}, Fourier transforms of the real and imaginary components of $a(t)$ and their sum. \textbf{b}, Fourier transforms of real and imaginary components of $a(t)$ simulated
    with a frequency shift of $\varepsilon = 2\pi \times \SI{30}{THz}$ and with frequencies corrected by $-\varepsilon$.}
    \label{fig:shft}
\end{figure}

The autocorrelation is a complex, Hermitian function, meaning $a(t) = a^*(-t)$ and thus $\mathrm{Re}\,a(t) = \mathrm{Re}\,a(-t)$ is even and $\mathrm{Im}\,a(t) = -\mathrm{Im}\,a(-t)$ is odd. The Fourier transform maintains parity and multiplies odd functions by $i$. Thus, $\mathcal{F}[\mathrm{Re}\,a(t)]$ is real and even, $\mathcal{F}[i\mathrm{Im}\,a(t)]$ is real and odd, and the spectrum $\mathcal{F}[a(t)]$ is the sum of the two, as shown in
Fig.~\ref{fig:shft}a. If all features of the spectrum appeared at strictly positive frequencies, it would mean that $\mathcal{F}[\mathrm{Re}\,a(t)] = \mathcal{F}[i\mathrm{Im}\,a(t)]$ for $\omega > 0$, and the two components would cancel for $\omega < 0$. Therefore, only one of the two would be necessary to produce the spectrum.

In cases where features of the spectrum (i.e., peaks and their linewidths) appear at $\omega \le 0$, we can introduce a frequency shift $\varepsilon$ such that all features appear at $\omega + \varepsilon > 0$.
This corresponds to subtracting a constant frequency term from the reference state, or
\begin{equation}
    \hat{H}'_{\varepsilon} = \hat{H}' - \hbar\varepsilon \ket{\mathrm{ref}}\bra{\mathrm{ref}}.
\end{equation}
Fig.~\ref{fig:shft}b shows the spectrum and its components when a frequency shift $\varepsilon$ is applied.
After the Fourier transform, the correct frequencies are restored by subtracting $\varepsilon$, i.e., translating the spectrum by $\varepsilon$ in frequency.
Because the real and imaginary components are equal to $\mathcal{F}[a(t)]/2$ for $\omega > -\varepsilon$, experimental measurement of $\mathrm{Re}\,a(t)$ is sufficient to obtain the spectrum.

Choosing $\varepsilon$ requires an estimate of the frequency and linewidth of the lowest-frequency peak in the spectrum. For some of Hamiltonians, such as QVC Hamiltonians with weak vibronic coupling near $Q = 0$, the lowest frequency can be estimated as the zero-point energy of the lowest excited electronic state. The linewidth can be estimated from the noise conditions of the simulator using previous experimental measurements. In the absence of a good estimate of the lowest frequency or its linewidth, low-resolution experiments (i.e., with a short propagation time) can be used to increase $\varepsilon$ until all features of the spectrum are well separated from $\omega = 0$.

\section{Experimental calibration}
\label{app:calibration}

\begin{figure}
    \centering
    \includegraphics[width=\columnwidth]{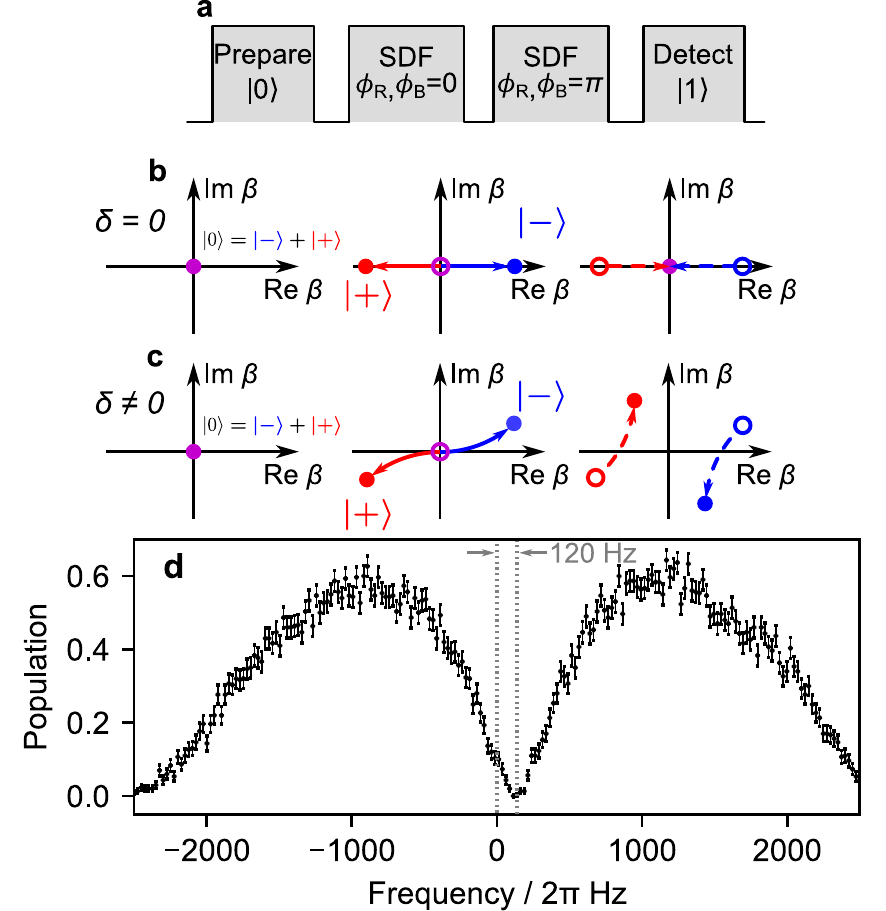}
    \caption{Calibration of symmetric frequency detuning. \textbf{a},~Pulse sequence for calibrating the symmetric detuning $\delta$ with respect to the motional sidebands. \textbf{b}, At resonance, the spin states move in straight trajectories and return to the origin, leading to final qubit state $\ket{0}$. \textbf{c}, In the presence of a frequency offset $\delta \neq 0$ during the first SDF pulse, the spin states $\ket{+}$ and $\ket{-}$ follow a circular trajectory, completing one loop in time $t = 2\pi/\delta$. For a pulse time $\tau<2\pi/\delta$, the spin states will be displaced by a small amount from the origin. The phase offset of $\phi_\mathrm{M} = \pi$ for the second SDF pulse again displaces along a similar trajectory, but with the centre of a circle displaced due to the first pulse. The residual displacement along the imaginary axis results in measuring significant population in the $\ket{1}$ state. \textbf{d}, The resonance condition is found by sweeping the offset detuning and observing where the population measurement is closest to 0. This data was collected by repeating the pulse sequence 300 times with an SDF pulse time of \SI{300}{\micro s}. The error bars correspond to uncertainty due to quantum projection noise.}
    \label{fig:detuning_scan}
\end{figure}

Operations with the trapped ion's degrees of freedom are driven by coherent laser interactions, and in this appendix we describe how the relevant laser parameters were calibrated.

In our experiment, three main laser interactions are used: carrier, and red- and blue-sideband transitions. Their Hamiltonians in the interaction picture are
\begin{align}
    \hat{H}_\mathrm{C}^I &= \hbar\frac{\Omega}{2}(\hat{\sigma}_{+}e^{i\phi_\mathrm{C}} + \mathrm{h.c.}), \label{eq:H_carrier}\\ 
    \hat{H}_\mathrm{R}^I &= \hbar\frac{\eta\Omega}{2}(\hat{\sigma}_{+} \hat{a} e^{-i(\delta t - \phi_\mathrm{R})} + \mathrm{h.c.}) \label{eq:H_R},\\
    \hat{H}_\mathrm{B}^I &= \hbar\frac{\eta\Omega}{2}(\hat{\sigma}_{+} \hat{a}^\dag e^{i(\delta t + \phi_\mathrm{B})} + \mathrm{h.c.}),\label{eq:H_B}
\end{align}
where the Rabi frequency $\Omega$ quantifies the coupling strength between the qubit states and the applied laser light, and $\hat{\sigma}_\pm = (\hat{\sigma}_x \mp i\hat{\sigma}_y)/2$. The  light-ion interaction imprints a phase relationship, which can be  controlled by the parameter $\phi_\mathrm{C}$, allowing for qubit state rotations around the $x$ or $y$ axis of the Bloch sphere. The sideband interactions in both Eq.~\ref{eq:H_R} and \ref{eq:H_B} are similar to the carrier interaction, but they also contain the bosonic ladder operators $\hat{a}$ and $\hat{a}^\dag$ for a single motional mode. Their interaction strength is scaled by the Lamb-Dicke parameter $\eta = \frac{2 \pi}{\lambda} \sqrt{\hbar/2 m \omega_x} = 0.084$, where $\lambda$ is the laser wavelength and $m$ is the ion's mass. $\eta$ is included in Eq.~\ref{eq:MSgate} via the sideband Rabi frequency $\Omega_\mathrm{S} = \eta\Omega/2$. Similar to the carrier interaction, the sideband interactions have associated phases $\phi_\mathrm{R}$ and $\phi_\mathrm{B}$. The detuning $\delta$ is a symmetric frequency offset from resonant motional sidebands.

\begin{figure}[t]
    \centering
    \includegraphics[width=\columnwidth]{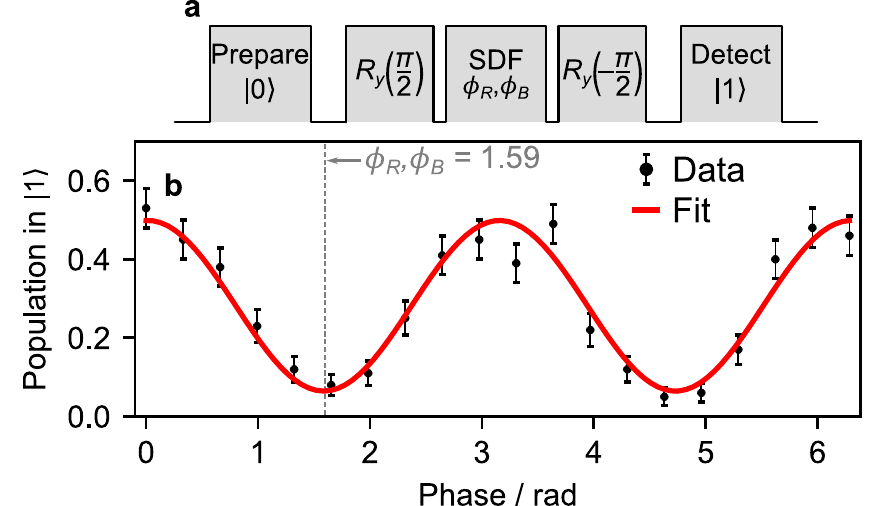}
    \caption{Bichromatic phase calibration. \textbf{a}, The pulse sequence used to calibrate the phase of the bichromatic field $\phi_\mathrm{S}$ with respect to the carrier phase $\phi_\mathrm{C}=0$ set by the first $\pi/2$ pulse. \textbf{b}, An example of the sine wave traced out by a full cycle of $\phi_\mathrm{S}$. The phase is varied and the operation of the SDF on its eigenstate $\hat{\sigma}_x$ is indicated by measurement of the qubit state having a near-zero population in $\ket{1}$. In this case, the phases $\phi_{\mathrm{R}}$ and $\phi_\mathrm{B}$ are set to 1.59 radians. The pulse sequence was repeated 100 times for each phase setting.}
    \label{fig:phase_scan}
\end{figure}

Simultaneously driving the red and blue sidebands implements a bichromatic pulse described by
\begin{align}
    \hat{H}_\mathrm{SB}^I &= \hat{H}_\mathrm{R}^I + \hat{H}_\mathrm{B}^I \nonumber \\
    &= \hbar\Omega_\mathrm{S}(\hat{\sigma}_+ e^{i\phi_\mathrm{S}} + \mathrm{h.c.})(\hat{a}^\dag e^{i(\delta t + \phi_\mathrm{M})} + \mathrm{h.c.}),
    \label{eq:HSB}
\end{align}
which depends on the spin phase $\phi_\mathrm{S} = (\phi_\mathrm{R} + \phi_\mathrm{B})/2$ and the motional phase $\phi_\mathrm{M} = (\phi_\mathrm{B} - \phi_\mathrm{R})/2$.
Experimentally, $\phi_\mathrm{S}$ and $\phi_\mathrm{M}$ can have an arbitrary, constant offset, and they are adjusted so that Eq.~\ref{eq:HSB} has equivalent phase relationships to Eq.~\ref{eq:MSgate}.
The bichromatic pulse is used to both initialise the ion's motional wavepacket into a displaced coherent state and to drive the simulated molecular time evolution. In the following, we describe how the physical parameters that make up the bichromatic fields are calibrated, namely their frequencies, amplitudes, phases, and durations.

Laser frequencies are calibrated as described in Fig.~\ref{fig:detuning_scan}, allowing the control of the detuning $\delta$ in Eq.~\ref{eq:HSB}.
The sequence consists of two SDF pulses of equal duration, with a $\pi$ phase shift between them~\cite{Milne:2021}. If the bichromatic fields are resonant with the motional sidebands, the spin and motion are disentangled after the SDF evolutions (see Fig.~\ref{fig:detuning_scan}b). However, in the presence of motional frequency offsets, the spin and motion remain entangled (see Fig.~\ref{fig:detuning_scan}c), leading to measurements of a partially mixed qubit state. Static frequency offsets are therefore calibrated by symmetrically varying the bichromatic fields' frequencies and measuring qubit population in the $\hat{\sigma}_z$ basis. Using this method, we are able to calibrate the frequencies of the bichromatic fields to within \SI{50}{Hz}. Furthermore, calibrations of the motional frequency are scheduled every 5 minutes to mitigate the effects of drift.

\begin{figure}
    \centering
    \includegraphics[width=\columnwidth]{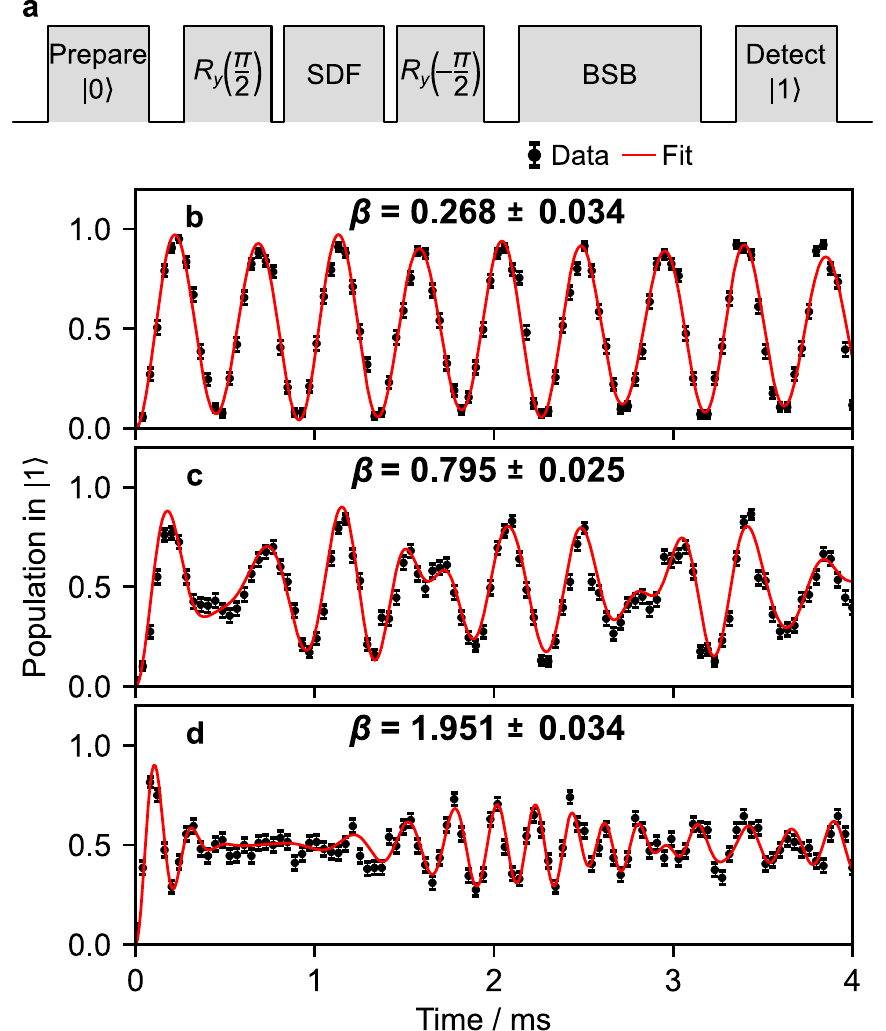}
    \caption{Calibrating the duration of the bichromatic pulses. \textbf{a}, The pulse sequence used to calibrate the displacement operation. Longer SDF pulses create larger displacements $\hat{D}(\beta)$. The displacement distance can be inferred from fitting the time evolution of a blue-sideband-driven population to Eq.~\ref{eq:BSB_fit}. Examples in \textbf{b}--\textbf{d} correspond to applying $\hat{D}(\beta)$ for $t$ = 0.05, 0.15 and 0.4 ms with $\Omega_\mathrm{S} = 2\pi \times \SI{0.850}{kHz}$. Each data point is an average of 200 repetitions.}
    \label{fig:laser_displacement}
\end{figure}

The amplitudes of the two tones of the bichromatic field, corresponding to the red and blue sidebands, are calibrated independently. An imbalance leads to an unwanted AC-Stark shift affecting the qubit frequency and residual coupling between the qubit state and the bosonic modes. We calibrate the amplitudes by measuring their Rabi frequencies from Rabi oscillations. The Rabi frequencies are equalised by adjusting the respective RF signal's amplitude. In practice, we find a difference of $<3\%$ between Rabi frequencies of two tones and negligible variation on the timescale of an experiment.

We use the pulse sequence depicted in Fig.~\ref{fig:phase_scan}a to calibrate the phases of the bichromatic fields such that the displacement operator enacted by the SDF interaction acts in the $\hat{\sigma}_x$ eigenbasis of the qubit. In the calibration, we set $\phi_\mathrm{R} = \phi_\mathrm{B}$ and vary them simultaneously. We find the phase for which the spin state is unchanged, indicating that the overall operation is acting on the $\hat{\sigma}_x$ eigenbasis.

Once the frequency, amplitude, and phase are calibrated with sufficient precision, the duration of the pulse can be calibrated to set the displacement to the desired value. The pulse sequence for the calibration is shown in Fig.~\ref{fig:laser_displacement}a. The magnitude $|\beta|$ of the displacement operator $\hat{D}(\beta)$ is determined by the Rabi frequency and the SDF pulse duration; for simplicity, we do not change the Rabi frequency and instead adjust only the pulse duration. After the displacement operation, the magnitude is estimated by observing the change in qubit state population after driving a blue-sideband transition~\cite{leibfried96}. The resulting spin probability follows
\begin{equation}
    P_{\ket{1}}(t) = \frac{1}{2} \Big(1 - e^{- \zeta t - \bar{n}}\sum_{k=0}^\infty\frac{\bar{n}^k}{k!}\cos(\Omega_k t)\Big),
    \label{eq:BSB_fit}
\end{equation}
where $\Omega_k = e^{-\eta^2/2}\eta\Omega L^1_k(\eta^2)\sqrt{k + 1}$ is the Rabi frequency of the blue sideband for Fock state $k$, $L^1_k(x)$ is the Laguerre polynomial in $x$ of order $k$, and the fitting parameter $\zeta$ introduces amplitude damping that might be present due to motional state decoherence. The observed oscillations are fitted to extract $|\beta|$ (see Fig.~\ref{fig:laser_displacement}b--d), allowing the bichromatic fields' duration to be varied to correct the displacement magnitude.

\end{document}